

\input phyzzx.tex
\unnumberedchapters
\pubnum{$\caps ROM2F - 93/33$}
\titlepage
\title{Sewing \ \ Constraints \ \ and \ \  Non \ - \ Orientable \ \
Open \ \  Strings}
\author{D. Fioravanti \ , \ \ G. Pradisi \ \ and \ \ A. Sagnotti}
\address{\it Dipartimento \ di \ Fisica\break
Universit\`a di Roma \ ``Tor Vergata''\break
I.N.F.N. \ -- \ Sezione di Roma \ ``Tor Vergata''\break
Via \ della \ Ricerca \ Scientifica , \ 1\break
00133 \  Roma \ \ ITALY}
\vskip .6 truein
\abstract
{\baselineskip=14pt
We extend to non-orientable surfaces previous work on sewing constraints
in Conformal Field Theory.  A new constraint, related to
the real projective plane, is described and is used to illustrate the
correspondence with a previous construction of open-string spectra.}
\endpage
\pagenumber=2
\chapter{Introduction}
\vskip 24pt
The main ingredient of Conformal Field Theory, central to all its applications
in String Theory and critical phenomena, is the extensive use of notions and
techniques from the classical theory of analytic
functions\Ref\bpz{A.A. Belavin, A.M. Polyakov and A.B. Zamolodchikov,
\nextline
{\sl Nucl. Phys.} {\bf B241} (1984) 333.}.  Thus,
amplitudes may be defined in terms of power series,
their full structure being determined by analytic
continuation.  The global aspects of the geometry manifest themselves in
a number of conditions that the expansion coefficients are to satisfy.
These are the ``sewing constraints''. In soluble models, they yield
simple algebraic conditions on the data of the conformal theory.

For amplitudes defined on closed orientable surfaces,
the case of interest for models of oriented closed strings,
only two sewing constraints are needed\Ref\son{H. Sonoda,
{\sl Nucl. Phys.} {\bf B311} (1988) 401, 417.}.  The first one is the
non-planar duality of the four-point amplitude on the sphere,
a condition on the OPE coefficients $C_{ijk}$ [\bpz],
while the second one is (essentially) the modular
invariance\Ref\car{J. Cardy, {\sl Nucl. Phys.} {\bf B240} (1986) 186.}
of the torus amplitude. One may then argue by induction in the number of
moduli that all amplitudes are consistent with the geometry.

Similar inductive arguments \Ref\carle{J. Cardy and
D.C. Lewellen, {\sl Phys. Lett.} {\bf B259} (1991) 274.}
\Ref\lew{D.C. Lewellen, {\sl Nucl. Phys.} {\bf B372} (1992) 654.}
show that, in order to define a conformal theory on surfaces with an
arbitrary number of holes, {\it four} additional constraints
need be satisfied.  Now there are both ``bulk''
and ``boundary'' operators (the conformal counterparts of closed-string and
open-string emission vertices), and the latter may also mediate changes of
boundary conditions.  Since in general there is some freedom in the choice
of boundary conditions\Ref\caran{J. Cardy, {\sl Nucl. Phys.}
{\bf B324} (1989) 581.}, the corresponding OPE coefficients
$C_{ijk}^{abc}$ contain additional labels that specify them. The
normalization coefficients of the
one-point functions in the presence of a boundary, $C_k^a$ [\carle],
and the one-point
functions of the identity in the presence of a boundary, $\alpha^a$,
are new data of the theory. The four additional constraints typify their
possible
choices.

This letter is aimed at extending the results to
non-orientable surfaces with an arbitrary number of boundaries, thus
relating refs. [\carle] and [\lew] to
the constructions of open-string descendants \Ref\as{A. Sagnotti,
{\it in} ``Non-Perturbative Quantum Field Theory'',
\nextline eds. G. Mack et al (Pergamon Press, 1988), p. 512.}
of closed-string models of refs.~ \Ref\bs{M. Bianchi and A. Sagnotti,
{\sl Phys. Lett.} {\bf B247} (1990) 517, \nextline{\sl Nucl. Phys.} {\bf B361}
(1991) 519.}, \ \Ref\bpsa{M. Bianchi, G. Pradisi and A. Sagnotti, {\sl Phys.
Lett.} {\bf B273} (1991) 389.} and \Ref\bpsb{M. Bianchi, G. Pradisi and A.
Sagnotti, {\sl Nucl. Phys.} {\bf B376} (1992) 365.}.
Standard results on the topology of Riemann
surfaces\Ref\ss{M. Schiffer and D. Spencer, ``Functionals of
Finite Riemann Surfaces'' \nextline (Princeton Univ. Press, 1954).},
first used in String Theory in ref. \ \Ref\alam{
V. Alessandrini, {\sl Nuovo Cimento} {\bf 2A} (1971) 321;\nextline
V. Alessandrini and D. Amati, {\sl Nuovo Cimento} {\bf 4A} (1971) 793.},
imply that their type is fully specified by the total
number of handles, the total
number of holes and the total number of crosscaps,
three crosscaps being topologically equivalent to one handle and
one crosscap. A crosscap may be introduced
by cutting a hole on the surface and by glueing to it a M\"obius
strip along its unique boundary. The crosscap is
our main interest here, because it provides a new building block of
the construction.  It may be modelled
as the real projective plane, {\it i.e.} the plane augmented by
a line at infinity.  Alternatively, it may be defined as an
orbifold of the sphere under the involution that identifies all pairs of
antipodal points.  Our results may be summarized as follows.
First of all, crosscaps allow for {\it three} new types of cuts (fig. 1).
Two of them join a crosscap to another crosscap or to a hole, and first
present themselves in Klein bottle and M\"obius amplitudes, where they
relate vacuum channels to ``twisted'' three-point functions.  The resulting
constraints were both implicitly taken into account in refs. [\bs], [\bpsa] and
[\bpsb].  On the other hand, the last cut first presents itself in the real
projective plane. It begins on one
crosscap and comes back to it after enclosing one puncture.  The resulting
constraint is new, and is related rather neatly to the fundamental group of the
twice
punctured real projective plane.  In the next Section we sketch the
extension of the inductive argument of ref. [\lew] and derive the crosscap
constraint.
In the last Section we illustrate the new constraint using the Ising model as
an
example.
\vskip 30pt
\chapter{The Real Projective Plane and the Crosscap Constraint}
\vskip 24pt
Fig. 1 displays the three new types of cuts allowed by the presence
of crosscaps.  They all connect pairs of identified points, since any line
surrounding
the cap may be deformed into a pair of these.  If the two
resulting ends are joined after enclosing at least one puncture, an allowed cut
results.  Alternatively, the two ends may terminate at a hole or at another
crosscap.
The last two settings require, respectively, a M\"obius strip and a Klein
bottle, and
were both implicitly taken into account in refs. [\bs], [\bpsa] and [\bpsb].
Let us
justify this statement for the M\"obius amplitude\foot{Following ref. [\lew],
we have
put one open-string puncture on the hole.}, whose vacuum channel exhibits the
propagation of a closed string between a hole (one of the building blocks of
ref.
[\lew]) and a crosscap. The crosscap amplitude for a closed-string puncture is
a  new building
block of the  construction.  It should be appreciated that the prescription for
the
vacuum channels of refs. [\bs], [\bpsa] and [\bpsb] endows the M\"obius
amplitude
with reflection coefficients that, for each sector of the spectrum, are
geometric means
of those for the other two vacuum channels.  This is precisely as demanded by
the
compatibility of the cuts.  The dual interpretation corresponds to the
ultraviolet
limit of the vacuum channel, and exhibits the open-string three-point function
with one
open-string ``twist''.  Though in principle an independent building block, this
amplitude is in fact proportional to the usual three-point function for each
set of
three physical states.  In a similar fashion, one may relate the Klein bottle
amplitude
of fig. 1
both to the propagation of a closed string between two crosscaps and to the
three-point
function on the sphere with one closed-string ``twist''.  These ``twist''
operations
are discussed in detail, within the old operator formalism, in ref.~
\Ref\sch{J.H.
Schwarz, {\sl Phys. Reports} {\bf 89} (1982) 223.}.  A reformulation in the
language
of conformal field theory will be presented elsewhere. We now turn to the
main topic of this letter, the first type of cut and the corresponding crosscap
constraint.

When working in the plane, the stereographic projection turns the
two-fold identification that defines the crosscap into the
anti-conformal involution (fig. 2)
$$
I(z) \ = \ - \ {1 \over \bar{z}} \qquad ,
\eqn\inv
$$
that has no fixed points and is compatible with the
$SU(2)$ group that transforms $z$ according to
$$
z \ \rightarrow \ {{a z + b} \over {- \bar{b} z + \bar{a}}} \qquad .
\eqn\sutwo
$$
This is the global symmetry of our problem.

In the more familiar case of a unit disk centered at the
origin, the involution lacks the ``minus'' sign in eq. \inv, and
the resulting line of fixed points is the unit circle $|z| = 1$.
The global symmetry, an $SU(1,1)$ subgroup of $SL(2,C)$, is
related by conjugation to the $SL(2,R)$ subgroup that obtains if the disk
is identified with the upper-half plane.  In the latter construction the
involution is $I(z) \ = \ \bar{z}$, and the
line of fixed points is the real axis.  The inversion present
in eq. \inv~plays an important role in the discussion that follows.

Despite its apparent simplicity, the crosscap is not topologically
trivial.  This non-orientable surface may be viewed as a disk where
two halves of the boundary are identified according to their (opposite)
orientations (fig. 3), and has a rather curious
feature: its fundamental group has a single non-trivial generator, $\alpha$,
that becomes contractible if ran along twice.  Indeed, if $\alpha$
were slipped across the right edge, it would emerge from the left edge with a
reverted orientation. The fundamental group of the crosscap is therefore
$Z_2$, the additive group of integers modulo two.  Amplitudes consistent with
it
require pairs of image punctures lying on pairs of
opposite rays through the origin.  For convenience, we confine one
puncture to the upper-half plane, though the unit circle would be
an equally good fundamental domain.

If $n$ punctures are present, any one of them may probe the fundamental
group of the surface with the other $n - 1$ punctures.  The simplest
non-trivial surface involves two punctures. Referring to fig. 4,
the puncture $A$ ``sees'' a fundamental group with two generators $\alpha$
and
$\beta$, where $\alpha^2 = \beta$. The crosscap constraint is the condition
that all
two-point amplitudes be single valued if one puncture, $A$ if you will, is
moved along
$\alpha$.  A recursive  argument along the lines of ref. [\lew] then shows that
amplitudes with arbitrary numbers of punctures are consistent as well.

In order to derive the crosscap constraint, let us begin by
defining the one-point function in the presence of a crosscap.  In the
notation and conventions of ref. [\lew], one would be tempted to write it
$$
{< \phi_k >}_a \ = \ {\Gamma_k^a \over { \bigl( z \ + \ {1 \over \bar{z}}
\bigr)^{2 h}}}	\qquad ,
\eqn\onepoint
$$
where the condition $h = \bar{h}$ is implicit and where $\Gamma_k^a$ are new
data
of the conformal theory associated to the crosscap.
Though natural in view of eq. \inv, this choice is not a convenient one, since
in the basis $( dz , dI(z) )$ an $( h,\bar{h} )$ differential has spurious
monodromies.  On the other hand, referring all differentials to the standard
basis $( dz , d \bar{z} )$, the resulting one-point function,
$$
{< \phi_k >}_a \ = \ {\Gamma_k^a \over {
\bigl( 1 \ + \ z \bar{z} \bigr)^{2 h}}}	\qquad ,
\eqn\onepoint
$$
is manifestly single-valued.

In order to define the two-point function, one needs the OPE of two
bulk operators,
$$
\phi_i (z) \ \phi_j (w) \ \sim \ \sum_k \ C_{ijk} \ (z - w)^{h_k -
h_i - h_j} \
(\bar{z} - \bar{w})^{\bar{h}_k - \bar{h}_i - \bar{h}_j}
\ \phi_k (w) \qquad .
\eqn\ope
$$
In the conventions defined above, this may be used to deduce the
OPE between one field and the image of the other. The result,
$$
\phi_i (z) \ I(\phi_j (w)) \ \sim \ \sum_k \ C_{ijk} \ \Gamma_k^a \
(1 + z \bar{w} )^{h_k - h_i - \bar{h}_j} \
(1 + \bar{z} w )^{\bar{h}_k - \bar{h}_i - h_j}
\ \phi_k (w) \quad ,
\eqn\opecap
$$
contains $\Gamma_k^a$ since, as in eq. \onepoint, the image of a field
need not coincide precisely with the field itself.

The two-point function contains two factors.  The first factor, $P$,
depends on the conformal weights and determines the behavior of the
amplitude under the residual projective
group.  In the conventions of ref. [\lew], and in the $( dz , d \bar{z} )$
basis for the differentials,
$$
\eqalign{P \ = \ ( z_1 - z_2 )^{r - h_1 - h_2} &{( \bar{z}_1 - \bar{z}_2
)}^{\bar{r} - \bar{h}_1 - \bar{h}_2} {( 1 + {| z_1 |}^2 )}^{r - h_1 -
\bar{h}_1} {( 1 + {| z_2 |}^2 )}^{r - h_2 - \bar{h}_2} \cr
&{( 1 + z_1 \bar{z}_2 )}^{r - h_1 - \bar{h}_2}
{( 1 + z_2 \bar{z}_1 )}^{r - h_2 - \bar{h}_1} \quad , \cr}
\eqn\pref
$$
where
$$
r \ = \ {1 \over 3} \ \bigl( h_1 + h_2 + \bar{h}_1 + \bar{h}_2 \bigr) \qquad .
\eqn\rfac
$$
The second factor, $Y$, is a function of the (real) cross-ratio
$$
\eta \ = \ {{{| z_1 - z_2 |}^2} \over { ( 1 + | z_1 |^2 ) ( 1 + | z_2 |^2 )}}
\qquad ,
\eqn\cross
$$
and is thus manifestly invariant under projective transformations.

The crosscap constraint results from the comparison of two distinct
limits of the two-point function. In the first case of fig. 4
$( \eta \rightarrow 0)$
two bulk operators approach one another, eq. \ope~
applies, and the limiting amplitude is the product of a three-point function
on the sphere and of the one-point function of eq. \onepoint. As in ref.
[\lew], $Y$ may then be related to the conformal blocks $F^k$, and the result
is
$$
Y ( \eta ) \ = \ \sum_k \ C_{ijk } \ \Gamma_k^a \ F^k ( \eta ) \qquad .
\eqn\firstlim
$$
In the second case of fig. 4 $( \eta \rightarrow 1 )$
one bulk operator approaches the image of the other
after moving along $\alpha$.  Since the quotient topology
allows one to effectively ``rotate'' the crosscap, the limiting amplitude,
determined by the OPE of eq. \opecap, is similar to the previous one.
Namely, it is again the product of a three-point function
on the sphere and of the one-point function of eq. \onepoint,
where one of the punctures is now replaced by its image. The
limiting behavior provides an independent determination of $Y$,
$$
Y ( \eta ) \ = \ \sum_k \ ( - 1 )^{h_i - \bar{h}_i + h_j - \bar{h}_j} \
C_{ijk} \ \Gamma_k^a \ F^k ( 1 - \eta )  \qquad ,
\eqn\seclim
$$
still linear in $\Gamma$, since in this case the limiting
one-point function involves an image originally in the upper-half plane,
and therefore has the
conventional normalization. The crosscap
constraint is the condition that the two definitions of eqs. \firstlim ~ and
\seclim ~ coincide.  In a rational model duality matrices relate the
different forms of the conformal blocks, and the crosscap constraint
becomes a linear equation for the $\Gamma_k^a$,
$$
\sum_k \ C_{ijk } \ \Gamma_k^a \ M {\biggl[ {{i \ \bar{j}} \atop {j \ \bar{i}}}
\biggr]}_{kp} \ = \ ( - 1 )^{h_i - \bar{h}_i + h_j - \bar{h}_j} \
C_{ijp } \ \Gamma_p^a \qquad .
\eqn\constr
$$
This is the main result of this work.  An inductive argument along the lines
of ref. [\lew] suggests that eq. \constr~completes the sewing constraints for
(rational) conformal models on arbitrary Riemann surfaces.

\vskip 30pt
\chapter{An  Example}
\vskip 24pt
In order to illustrate the content of eq. \constr, let us consider the
``open-string descendants'' of the Ising model [\bpsa].  In this case,
the spectrum of bulk operators is obtained combining the
familiar torus partition function
$$
T \ = \ {| \chi_0 |}^2 \ + \ {| \chi_{1/2} |}^2 \ + \  {| \chi_{1/16} |}^2
\qquad ,
\eqn\tor
$$
properly halved to account for the projection, with the Klein-bottle
contribution
$$
K \ = \ {1 \over 2 } \ \bigl( \ \chi_0 \ + \ \chi_{1/2} \ +
\ \chi_{1/16} \ \bigr) \qquad .
\eqn\klein
$$
As in the usual case, the bulk spectrum contains the primary
fields $1$, $\epsilon$ and $\sigma$,
of dimensions $(0,0)$, $(1/2,1/2)$ and $(1/16,1/16)$, but their Verma modules
are now truncated according to the ``parameter-space'' projection.
In addition, the
spectrum of boundary operators includes three types of primary fields, of
dimensions $0$, $1/2$ and $1/16$, with an associated pattern of Chan-Paton
charges \Ref\charges{J.E. Paton and H.M. Chan, {\sl Nucl. Phys.} {\bf B10}
(1969) 516;\nextline
J.H. Schwarz, {\it in} "Current Problems in Particle Theory", \nextline Proc.
Johns
Hopkins {\bf 6} (Florence, 1982);\nextline
N. Marcus and A. Sagnotti, {\sl Phys. Lett.} {\bf 119B} (1982) 97,
{\bf 188B} (1987) 58.}
determined by combining the annulus
$$ \eqalign{A \ = \ {1 \over 2} \ &\bigl(
n_0^2 \ + \ n_{1/2}^2 \ + n_{1/16}^2 \bigr) \ \chi_0 \cr &+ \ \bigl( n_0 \
n_{1/2} \ + \ {1 \over 2} \ n_{1/16}^2 \bigr) \ \chi_{1/2} \  + \ \bigl( n_0 \
n_{1/16} \ + \ n_{1/2} \ n_{1/16} \bigr) \ \chi_{1/16} \cr}
\eqn\ann
$$
and M\"obius  partition functions
$$
M \ = \ \pm \ {1 \over 2 } \ \bigl[ \bigl( n_0 \ + \ n_{1/2}
\bigr) \ \chi_0  \ + \ n_{1/16} \ \chi_{1/2} \bigr] \qquad ,
\eqn\mob
$$
where the overall factors enforce the projection in the Chan-Paton
charge space.  The three types of charges, of multiplicities $n_0$,
$n_{1/2}$ and $n_{1/16}$, correspond to as many types of boundaries [\caran],
and the charge assignments are manifestly compatible with the
factorization of amplitudes.

Making use of the matrix
$$
S  \ = \ {1 \over 2} \ \pmatrix{1&1&\sqrt{2}\cr
1&1&- \sqrt{2}\cr
\sqrt{2}&- \sqrt{2}&0\cr}
\eqn\smatrix
$$
in eqs. \klein~and \ann, and of the matrix $P=T^{1/2} \ S \ T^2 \ S \ T^{1/2}$
in eqs. \mob, one may turn them into the corresponding
vacuum-channel contributions,
$$
\tilde{K} \ = \ {1 \over 4} \ \bigl[ ( 2 + \sqrt{2} ) \chi_0 \ + \
( 2 - \sqrt{2} ) \chi_{1/2} \bigr]		\qquad \qquad , \qquad \qquad
\eqn\ktilde
$$
$$
\eqalign{ \tilde{A} \ = \ {1 \over 4} \ &\bigl[ {( n_0 + n_{1/2} + \sqrt{2}
\ n_{1/16} )}^2
\ \chi_0 \cr &+ \ {( n_0 + n_{1/2} - \sqrt{2} \ n_{1/16} )}^2 \ \chi_{1/2} \
 + \
\sqrt{2} \ {( n_0 - n_{1/2} )}^2 \bigr] \cr} \qquad ,
\eqn\atilde
$$
and
$$
\tilde{M} \ = \ \pm \ \biggl[ cos \bigl( {\pi \over 8} \bigr) \
( n_0 + n_{1/2} + \sqrt{2} \ n_{1/16} ) \ \chi_0 \ + \
sin \bigl( {\pi \over 8} \bigr) \
( n_0 + n_{1/2} - \sqrt{2} \ n_{1/16} ) \ \chi_{1/2} \biggr]  \quad .
\eqn\mtilde
$$
The consistency of this construction rests on the relation
between eq. \mtilde~and the other vacuum amplitudes of eqs.
\ktilde ~ and \atilde.  Since all may be obtained by sewing bulk one-point
amplitudes for boundaries and crosscaps, the
coefficients in eq. \mtilde~ should be (and indeed are) geometric means of
those in the other two  amplitudes\foot{As in ref. [\bpsa], we have
inserted in eq. \mtilde~the combinatoric factor that a proper
measure over the moduli would generate.}.
Actually, {\it all} the basic bulk one-point functions may be
extracted from
the vacuum amplitudes of eqs. \ktilde, \atilde ~ and \mtilde.  Thus,
apart from a combinatoric factor $1/2$ related to the parameter-space
orbifold construction, the square roots of the coefficients
in eq. \atilde ~ {\it are} the normalizations of the boundary amplitudes.
With a slight abuse of notation\foot{We are
omitting the additional factors ${(2 Imz)}^{- 2h}$.}, they are
$$
\eqalign{
&< {\bf 1} >_{b{\bf 1}} \ = \ {1 \over \sqrt{2}}	  \qquad
< {\bf 1} >_{b{\epsilon}} \ = \ {1 \over \sqrt{2}}	\qquad
< {\bf 1} >_{b{\sigma}} \ =   1  \qquad , \cr
&< \epsilon >_{b{\bf 1}} \ = \ 	{1 \over \sqrt{2}}   \qquad
< \epsilon >_{b{\epsilon}} \ = \ 	{1 \over \sqrt{2}} \qquad
< \epsilon >_{b{\sigma}} \ = \ - 1    \qquad , \cr
&< \sigma >_{b{\bf 1}} \ = \ 	 {1 \over {\root 4 \of 2}}  \qquad
< \sigma >_{b{\epsilon}} \ = \ 	- \ {1 \over {\root 4 \of 2}} \qquad
< \sigma >_{b{\sigma}} \ = \ 0   \qquad .\cr 	 }
\eqn\onepointb
$$
The ratios of these normalizations may be recovered from the
sewing constraints for the disk geometry [\lew].

A similar argument relates the vacuum Klein-bottle amplitude to a
single type of crosscap one-point functions,
$$
< {\bf 1} >_c \ = \ \sqrt{{2 + \sqrt{2}} \over 2}	  \qquad
< \epsilon >_c \ = \ \sqrt{{2 - \sqrt{2}} \over 2}	\qquad
< \sigma >_c \ =   0  \qquad .
\eqn\onepointc
$$
Clearly, the crosscap constraint {\it does not} fix
the absolute normalizations in eq. \onepointc. Moreover,
the Klein-bottle vacuum channel defines a single one-point amplitude for
each primary field, and this amplitude may be regarded as a definition of the
corresponding crosscap state.  This is to be contrasted to the case of
boundary states, where a preferred basis is selected by the fusion rules
involving boundary fields (or, alternatively, by the Chan-Paton
charge assignments). Still, eq. \constr ~ allows {\it only} the
parameter-space projection of eq. \klein, since it forces
$< \sigma >_c$ to vanish, while fixing the ratio of the other amplitudes.
Most of the relevant data, OPE coefficients $C_{ijk}$ and duality matrices,
may be found in ref. [\lew].
\vskip 30pt
\chapter{Acknowledgements}
\vskip 24pt
D. F. would like to thank the I.N.F.N. Gran Sasso Laboratory,
and in particular Prof. P. Monacelli, for granting him partial support
during the course of this research. A. S. would like
to thank I.C.T.P. and S.I.S.S.A., and in particular Prof. P. Fr\'e, for
the hospitality
extended to him while this work was being completed.
This work was supported in part by E.C. contract $SCI-0394-C$.
\vskip 30pt
\chapter{Figure Captions}
\vskip 24pt
\item{\bf Figure \ 1.} New types of cuts and the simplest settings that allow
them.
\item{\bf Figure \ 2.} A puncture and its image under the involution of eq.
(2.1).
\item{\bf Figure \ 3.} The generators of the fundamental group of the
crosscap.
\item{\bf Figure \ 4.} Factorizations of the two-point amplitude.
\endpage
\refout
\end